\documentclass[final,5p,times,twocolumn]{elsarticle}

\usepackage[T1]{fontenc}
\usepackage[utf8]{inputenc}
\usepackage{physics}
\usepackage{amsmath, amssymb}
\usepackage{tabularx,booktabs,array,dcolumn}
\usepackage{setspace}
\usepackage{vmargin}
\usepackage{xcolor}
\usepackage{graphicx,psfrag,subfigure}

\usepackage{float}
\usepackage[all]{xy}
\usepackage{verbatim}
\usepackage{dsfont}

\newcommand{\bA}{\boldsymbol{A}}
\newcommand{\ubA}{\underline{\boldsymbol{A}}}

\newcommand{\bp}{\boldsymbol{p}}

\newcommand{\bs}{\boldsymbol{s}}

\newcommand{\br}{{\boldsymbol{r}}}

\usepackage{dcolumn}%
\newcolumntype{.}{D{.}{.}{8}}
\DeclareUnicodeCharacter{2009}{\,}
\usepackage{bm}
\usepackage{mathtools}
\usepackage{physics}
\usepackage{comment}

\newcommand{\bos}[1]{\boldsymbol{#1}}

\newcommand{\mr}[1]{\mathrm{#1}}

\def\eem{\mr{e}}
\def\nb{N_\text{b}} 

\def\Eh{\text{E}_\text{h}}
\def\muEh{\mu\text{E}_\text{h}}
\def\nbas{{N_\text{b}}}

\def\nnuc{{N_\text{nuc}}}
\def\np{{n_\text{p}}}

\def\Rhh{{R_{\text{H}_2\cdots \text{H}}}}
\def\Htwo{{H$_2$}}
\def\Hthree{H$_3$}
\def\hmol{^{\text{H}_2}}
\def\hatom{^{\text{H}}}
\def\init{{\text{I}}}
\def\som{Supplementary Material}

\def\start{^{(0)}}

\usepackage[unicode]{hyperref}
\hypersetup{
   unicode=true,          
   plainpages=false,
   colorlinks=true,       
   citecolor=blue,        
}
\usepackage{soul}

\journal{Chemical Physics Letters}

\begin{document}

\begin{frontmatter}



\title{Benchmark potential energy curve for collinear \Hthree }


\author[inst1]{Dávid Ferenc}
\author[inst1]{Edit Mátyus\corref{cor1}}
\ead{edit.matyus@ttk.elte.hu}
\cortext[cor1]{Corresponding author}

\affiliation[inst1]{organization={ELTE, E\"otv\"os Lor\'and University, Institute of Chemistry},
            addressline={P\'azm\'any P\'eter s\'et\'any 1/A}, 
            city={Budapest},
            postcode={H-1117}, 
            country={Hungary}}

\begin{abstract}
A benchmark-quality potential energy curve is reported for the \Hthree{} system in collinear nuclear configurations. 
The electronic Schrödinger equation is solved using explicitly correlated Gaussian (ECG) basis functions using 
an optimized fragment initialization technique that significantly reduces the computational cost.
As a result, the computed energies improve upon recent orbital-based and ECG computations. 
Starting from a well-converged basis set, a potential energy curve with an estimated sub-parts-per-billion precision is generated for a series of nuclear configurations using an efficient ECG rescaling approach.
\end{abstract}


\end{frontmatter}


\section{Introduction}
\label{sec:intro}
The simplest chemical reaction \Htwo{} + H $\rightarrow$ H + \Htwo{}---and it's isotopologues---is possibly one of the most exhaustively studied chemical processes \cite{AoBaHe2005}. 
Furthermore, the \Hthree{} system has qualitatively interesting features: a shallow van-der-Waals minimum for collinear nuclear structures and a conical intersection for equilateral triangular configurations. 
These features impose challenges when investigating the quantum dynamics of the system and require a high-level description of the electronic structure. 
The first potential energy surface (PES) for collinear \Hthree{} was obtained by Liu in 1973~\cite{Liu1973}. Since then, several full-dimensional surfaces have been published \cite{LSTH1,LSTH2,DMBE,BKMP,BKMP2,wu1999,DiAn1994} and refined \cite{BlombergLiu1985,BaStLaPa1990,PaBaStEu1993,DiAn1992,MiGaBrPe1999,RiAn2003,Hong-Xin2005} using increasingly accurate quantum chemical methods. More recently, a multireference configuration interaction (MRCI) PES was developed, using a hierarchy of correlation consistent basis sets followed by extrapolation to the complete basis set (CBS) limit \cite{MiGaPe-H3PES} with an estimated $\mu\text{E}_\text{h}$ level of  precision.
This complete configuration interaction (CCI) surface has been the most accurate full-dimensional PES of \Hthree{}, and it was used to resolve long-standing discrepancy of experimental and theoretical thermal rate constants \cite{Mielke2003}. 

The first computation for this system using explicitly correlated Gaussian (ECG) basis functions was performed by Cafiero and Adamowicz \cite{CafAdam2001}. They determined the stationary points of the PES by the simultaneous minimization of the energy with respect to both the nonlinear parameters of the basis functions and the nuclear configuration using analytic gradients. Nevertheless, using only 64 basis functions, they obtained an energy, $-1.673\ 467~\Eh$, which is above the dissociation threshold,  $E(\text{\Htwo{}})+E(\text{H})=-1.674\ 475\ 714~\Eh$.

In later work, Pavanello, Tung, and Adamowicz carried out methodological developments to improve the convergence of the ECG wave function and energy, and to reduce the computational cost for polyatomic, \emph{i.e.,} H$_3^+$ and \Hthree{}, systems. Their efforts resulted in the most precise non-relativistic energy for \Hthree{}, so far, near the equilibrium structure~\cite{PaTuAd2009}. 

The aim of the present letter is to explore and take the achievable precision further for H$_3$, a simple prototype for poly-electronic and poly-atomic molecular systems, using explicitly correlated Gaussian functions.

\section{Method \label{sec:Method}}
The Schrödinger equation (in Hartree atomic units) with $\nnuc$ nuclei clamped at the $\bos{R}$ configuration and $\np$ electrons,
\begin{align}
    H\psi(\br;\bos{R}) = E(\bos{R}) \psi(\br;\bos{R})
\end{align}
\begin{align}
    H = \frac{1}{2} \sum_{i=1}^\np \bp_i^2 - \sum_{a=1}^\nnuc \sum_{i=1}^\np \frac{Z_a}{r_{ia}} &+ \sum_{i<j}^\np \frac{1}{r_{ij}} + \sum_{a<b}^\nnuc \frac{Z_aZ_b}{R_{ab}} \; ,
\end{align}
is solved for the ground state of \Hthree{} using a set of floating ECG basis functions,
\begin{align}
    \psi(\br;\bos{R}) 
    &= 
    \mathcal{A} \sum_{n=1}^\nbas 
      c_n  \phi_n(\br;\bA_n,\bs_n) \chi_n(\vartheta) \\
    \phi_n(\br;\bA_n,\bs_n) 
    &= 
    \exp\left[ -(\br-\bs_n)^\text{T} \ubA_n (\br-\bs_n) \right] \; ,
\end{align}
where $\ubA_n=\bA_n\otimes I_3$, 
$\bA_n\in\mathbb{R}^{\np\times \np}$ is the exponent matrix and $\br,\bs \in \mathbb{R}^{3\np}$ are the coordinate vectors of the electrons and the Gaussian centers, respectively. 
$\mathcal{A}$ is the anti-symmetrization operator, and $\bA$ is parameterized in the $\bA = \bos{L}^T\bos{L}$ Cholesky-form, with an $\bos{L}$ lower-triangular matrix, to ensure positive definiteness of $\bA$ and square integrability of the basis functions. 
The A$_1$ symmetry (in the $C_{\infty v}$ point group) of the ground-state wave function is realized by constraining the Gaussian centers to the $z$ axis. 

The $\chi_n(\vartheta)$  three-particle spin function corresponding to the doublet multiplicity of the ground-state is obtained as a linear combination of the two possible couplings of the elementary, one-electron spin functions $\sigma(i)_{\frac{1}{2},\pm\frac{1}{2}}$ to a doublet state \cite{SuzukiVarge1998},
\begin{align}
    \chi_n &= 
    d_{n_1} \left[\left[ \sigma(1)_{\frac{1}{2}} \sigma(2)_{\frac{1}{2}} \right]_{1,0} \sigma(3)_{\frac{1}{2}} \right]_{\frac{1}{2},\frac{1}{2}}\nonumber  \\
    &\ 
    + d_{n_2} \left[\left[ \sigma(1)_{\frac{1}{2}} \sigma(2)_{\frac{1}{2}} \right]_{0,0} \sigma(3)_{\frac{1}{2}}\right]_{\frac{1}{2},\frac{1}{2}} \; ,
    \label{eq:spincoupling}
\end{align}
where the square brackets denote angular momentum coupling, using the Clebsch--Gordan coefficients $\braket{j_1,m_{j_1},j_2,m_{j_2}}{J,M_J}$. For example, coupling two spin-1/2 particles to a singlet function is labelled as
\begin{align}
    &\left[ \sigma(1)_{\frac{1}{2}} \sigma(2)_{\frac{1}{2}} \right]_{0,0} \nonumber \\ 
    &= \braket{\frac{1}{2},\frac{1}{2},\frac{1}{2},-\frac{1}{2}}{0,0} \sigma(1)_{\frac{1}{2},\frac{1}{2}}\sigma(2)_{\frac{1}{2},-\frac{1}{2}} \nonumber \\
    &+\braket{\frac{1}{2},-\frac{1}{2},\frac{1}{2},\frac{1}{2}}{0,0} \sigma(1)_{\frac{1}{2},-\frac{1}{2}}\sigma(2)_{\frac{1}{2},\frac{1}{2}} \nonumber \\
    &=\frac{1}{\sqrt{2}}\left( \ket{\uparrow\downarrow} - \ket{\downarrow\uparrow} \right)   \; .
\end{align}
Considering the normalization condition as well, the doublet three-electron spin functions can be parameterized by a single $\vartheta_n$ parameter as 
\begin{align}
  d_{n_1} = \sin \vartheta_n
  \quad\text{and}\quad
  d_{n_2}=\cos\vartheta_n \; , 
  \label{eq:theta}
\end{align}
and $\vartheta_n$ is optimized together with the nonlinear parameters of the basis set.

\subsection{Optimized fragment initialization \label{sec:init}}
The initial basis function parameters are usually generated in a pseudo-random manner, retaining those functions from a trial set that provide the lowest energy expectation value. This generation procedure is followed by extensive refinement of the parameterization based on the variational principle \cite{SuzukiVarge1998}. 
By increasing the number of electrons, the dimensionality of the parameter space, and hence, the optimization cost increases. 
To keep the computational cost low, it is useful to consider that the interaction between the electrons of the hydrogen molecule and the electron of the hydrogen atom is weak in the van-der-Waals well or if the two `fragments' are not too close, in general. If the interaction is not too strong, then a $\psi_\init$ initial approximation for the wave function can be written as the product of the wave functions optimized for the `fragments' (atom and molecule for the present example):
\begin{align}
    \psi_\init^\text{\Hthree} (\br_1,\br_2,\br_3) &= \psi^\text{\Htwo} (\br_1,\br_2) \psi^\text{H} (\br_3) \nonumber \\
    &=
    \sum_{k,l} c_k c_l \phi_k^{\text{H}_2}(\br_1,\br_2) \phi_l^{\text{H}} (\br_3) \; ,
    \label{eq:psi0}
\end{align}
which corresponds to an initial parameterization of the three-electron basis set with
\begin{align}
  \bA_{kl}^\init
  =
  \left(%
    \begin{array}{@{}c@{\ }c@{}}
      \bos{A}\hmol_k & 0 \\
      0  &  A\hatom_l \\
    \end{array}
  \right) \; ,
\end{align}
and the 3-electron $\bs$ vectors include the $\bs$ vectors shifted according to the configuration of the `fragments' in H$_3$:
\begin{align}
  \bs_{kl}^\init
  =
  \left(%
    \begin{array}{@{}c@{}}
      \bs\hmol_k+\bos{R}\hmol_\text{CM} \\[0.1cm]
      \bs\hatom_l+\bos{R}\hatom
    \end{array}
  \right) \; ,
\end{align}
where $\bos{R}\hmol_\text{CM}$ is the center of mass of the protons in H$_2$.

This procedure is reminiscent of the monomer contraction method that was first introduced in Ref.~\cite{CeKoPaSz2005} for the helium dimer, although there are a few differences.
First, we use the fragment (or monomer) basis set only to initialize the many(three)-electron basis, and we run repeated refinement cycles \cite{MaRe2012,Matyus2019} using the Powell method \cite{Po2004} for this initial basis.
Second, retaining the full direct-product basis optimized for H$_2$ and separately for H would be computationally very demanding, so instead, we truncate the direct-product basis according to the following strategy.

\begin{figure}
    \centering
    \includegraphics[width=\linewidth]{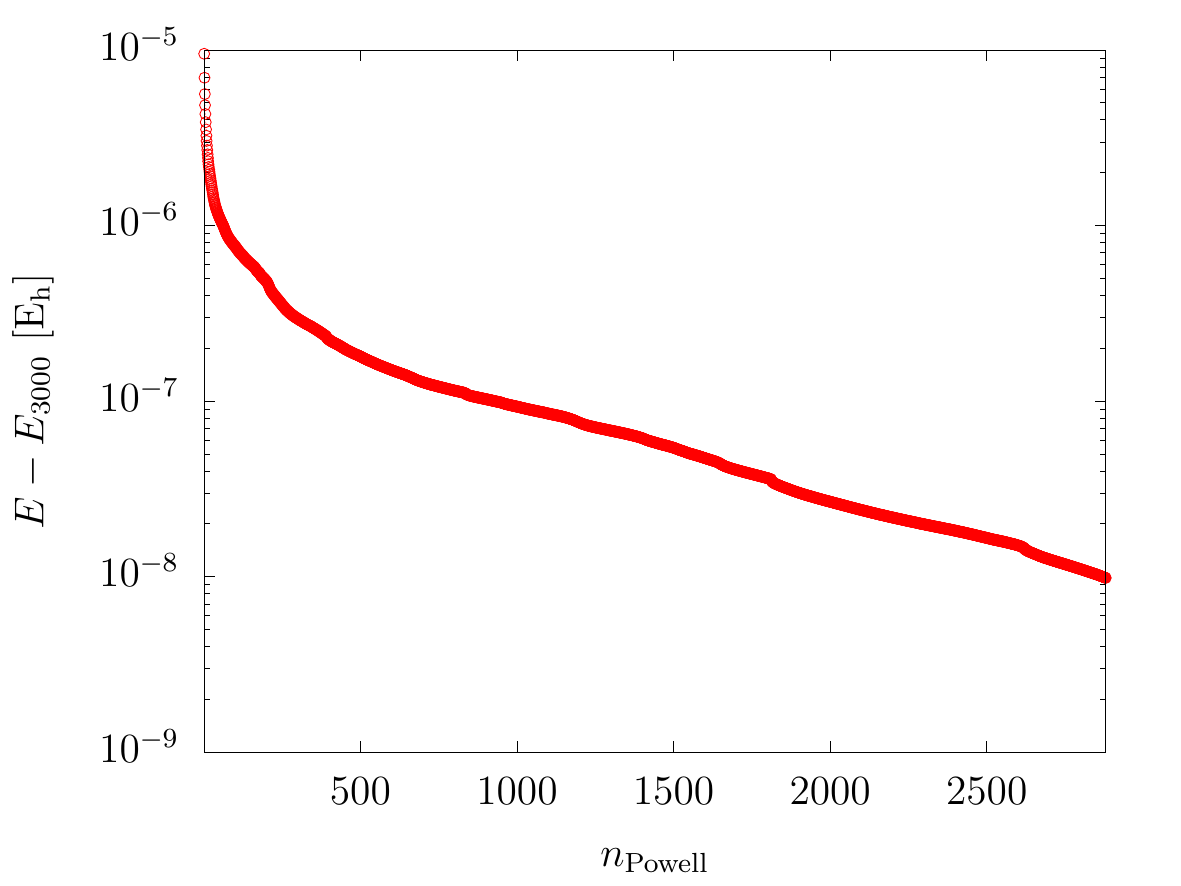}
    \caption{%
      Convergence of the ground-state energy of H$_2\cdots$H during the course of the Powell refinement cycles ($n_\text{Powell}$) of $N_\text{b}=1200$ basis functions initialized using basis functions optimized for the fragments, Eq.~(\ref{eq:init}).
      $R_\text{\Htwo} = 1.4$~bohr and $\Rhh = 6.442$~bohr, 
      $E_\text{3000} = -1.674\ 561\ 687~\Eh$.
      (See also Table~\ref{tab:convergence}.)
      }
    \label{fig:conv}
\end{figure}

The ground-state wave function of the \Htwo{} molecule was expanded over 1200 ECG functions, yielding $-1.174\ 475\ 714\ \Eh$ for the ground state energy, which---compared to the most accurate value obtained by Pachucki $-1.174\ 475\ 714\ 220\ 443 4(5)\ \Eh$ \cite{Pachucki2010}---is converged to a fraction of a n$\Eh$. 
The wave function of the hydrogen atom was represented with 10 optimized Gaussian functions, resulting in $-0.499\ 999\ 332\ \Eh$ (in comparison with the exact value, $-0.5\ \Eh$) ground-state energy. 
Inclusion of all possible combinations of the H$_2$ and H basis functions would result in a gigantic, $12\ 000$-term expansion. Such a long expansion would be prohibitively expensive to extensively optimize (refine), and it is unnecessary to have so many functions for the reaching a $1:10^9$ (ppb) precision.
To reduce the direct-product basis, 
it would be possible to perform competitive selection over the large basis space
or to order (and then truncate) the basis functions based on their importance in lowering the energy \cite{SuzukiVarge1998}. 
In the present work, we used a very simple construct that does not require any computation: 
we have generated a set of 1200 functions by appending each H$_2$ basis function from the 1200 set with a single H function. Out of the 10 H functions, we have picked one based on the basis index, \emph{i.e., } 
\begin{align}
  &\left\lbrace \phi\hmol_{10n+i}\phi\hatom_{i}; n=0,1,\ldots,119, i=1,2,\ldots,10 \right\rbrace \nonumber \\
  &=
  \left\lbrace %
    \phi\hmol_{1} \phi\hatom_1, 
    \phi\hmol_{2} \phi\hatom_2, 
    \ldots, 
    \phi\hmol_{10} \phi\hatom_{10}, \right.\nonumber \\
  &\quad 
  \phi\hmol_{11} \phi\hatom_1, 
  \phi\hmol_{12} \phi\hatom_2, 
  \ldots, 
  \phi\hmol_{20} \phi\hatom_{10}, \nonumber \\
  &\quad \ldots \nonumber \\
  &\quad \left. 
  \phi\hmol_{1191} \phi\hatom_1, 
  \phi\hmol_{1192} \phi\hatom_2, 
  \ldots, 
  \phi\hmol_{1200} \phi\hatom_{10} \right\rbrace \; .
  \label{eq:init}
\end{align}
The spin basis functions defined in Eq.~\eqref{eq:spincoupling}, were initialized by coupling the two electrons initially localized on the \Htwo{} fragment to a singlet state, \emph{i.e.,} $d_{n_1}=0$ and $d_{n_2}=1$ corresponding to $\vartheta_n=0$ ($n=1,2,\ldots,1200$) in Eqs.~\eqref{eq:spincoupling}--\eqref{eq:theta}. All non-linear parameters, including $\vartheta_n$, of the initial basis set were excessively optimized in repeated refinement cycles (Fig.~\ref{fig:conv}). The optimized fragment-based initialization of the basis set, described in this section, allowed saving several weeks (months) of computer time in comparison with Ref.~\cite{PaAd2009} (see also Sec.~\ref{sec:res}).

\subsection{Gaussian-center scaling \label{sec:scaling}}
Independent variational optimization of the basis set at may points along the PEC (or over the PES) would make the computations very computationally intensive. Ko\l os and Wolniewicz \cite{KolosWolniewicz1964} noted already in 1964 that 
for a sufficiently large basis set, the $\bA_k$ exponents are insensitive to small displacements of the nuclear coordinates. In 1997, Cencek and Kutzelnigg proposed a scaling technique to generate a good initial ECG (re)parameterization for the electronic basis set of diatomics upon small nuclear displacements \cite{CeKu1997}. They noted that their approach can be generalized beyond diatomics. Pavanello and Adamowicz implemented rescaling the ECG centers (to have a good starting basis set) of H$_3^+$ upon small nuclear displacements to generate a series of points to represent the 3D PES \cite{PaAd2009,PaTuLeAd2009,AdPa2012,H3pPES2012}. 
Upon a small $\Delta\bos{R}_a$ displacement of the coordinates of the $a$th nucleus,
\begin{align}
  \bos{R}_a' &= \bos{R}_a + \Delta \bos{R}_a \; ,
\end{align}
the $\bs_i\in\mathbb{R}^3$ ECG centers corresponding to the $i$th electron were transformed as
\begin{align}
  \bs_i' &= \bs_i + \Delta \bs_i  \; ,
\end{align}
where $\Delta\bs_i$ is expressed as a function of the $\Delta \bos{R}_a$ nuclear displacement,
\begin{align}
    \Delta \bs_i 
    &= 
    \frac{1}{W_i} 
    \sum_{a=1}^\nnuc 
    w_{ia}\ \Delta \bos{R}_a \; 
    \label{eq:rescal}
\end{align}
with $W_i=\sum_{a=1}^\nnuc w_{ia}$. The $w_{ia}$ `weight' is a function constructed based on simple arguments. 
It is chosen to be the distance of the $\bs_i$ center and the $a$th nucleus, $|\bs_i-\bos{R}_a|$
and it is expected to have good limiting properties. 
First, it must vanish if the $\bs_i$ center is very (infinitely) far from the displaced nucleus, $\lim_{|\bs_i-\bos{R}_a|\rightarrow\infty}w_{ia}=0$. 
Second, the closer the $\bs_i$ center to the $\bos{R}_a$ nucleus position, the $\Delta \bos{R}_{ia}$ displacement has a larger contribution, \emph{i.e.,} larger $w_{ia}$ weight, to the $\Delta\bos{s}_i$ change.

These conditions allow several possible choices for the weight function. For example, Coulomb-like weights were used in Ref.~\cite{PaAd2009} 
\begin{align}
    w_{ia}^\text{C} = \frac{1}{|\bs_i-\bos{R}_a|} \; .
    \label{eq:coul}
\end{align}
After some experimentation with different possible functions, and inspired by the picture that the weight function can be intuitively defined as if there was some attraction between the centers and the nuclear positions by a central field, a Yukawa-like weight function appears to be a good choice 
\begin{align}
    w_{ia}^\text{Y} = \frac{\eem^{-\mu |\bs_i-\bos{R}_a|}}{|\bs_i-\bos{R}_a|} \; ,
    \label{eq:yuk}
\end{align}
where the parameter $\mu\in\mathbb{R}^+$ was set to unity in this work.
For small nuclear displacements, a parameterization rescaled with Yukawa weights (with $\mu=1$) provided an energy lower than rescaling with Coulomb weights, Eq.~(\ref{eq:coul}). 

The rescaling technique with the Yukawa weight function was used to generate the PEC
corresponding to the H atom approaching the \Htwo{} molecule with a proton-proton distance fixed at $R_{\text{H}_2}=1.4$~bohr. The $\Rhh$ distance of the hydrogen atom was measured from the center of mass of the \Htwo{} fragment. 
The starting value was $\Rhh =6.442$~bohr, for which an initial basis set was generated using the optimized fragment initialization (Sec.~\ref{sec:init})
and the representation was improved through several Powell refinement \cite{Po2004} cycles of the non-linear parameters (Fig.~\ref{fig:conv}). Then, initial basis sets were generated by making small $\Delta\Rhh=\pm0.1$~bohr displacements, rescaling the centers according to Eq.~(\ref{eq:rescal}) with Yukawa weights, Eq.~(\ref{eq:yuk}), followed by 5 entire basis refinement cycles (that took 4 hours) before the next step was taken along the series of the nuclear configurations (the positive and the negative displacement series were run in parallel).
All computations have been carried out using the QUANTEN computer program \cite{Matyus2019,FeMa2019HH,FeMa2019EF,FeKoMa2020}.

\begin{figure}
\includegraphics[width=\linewidth]{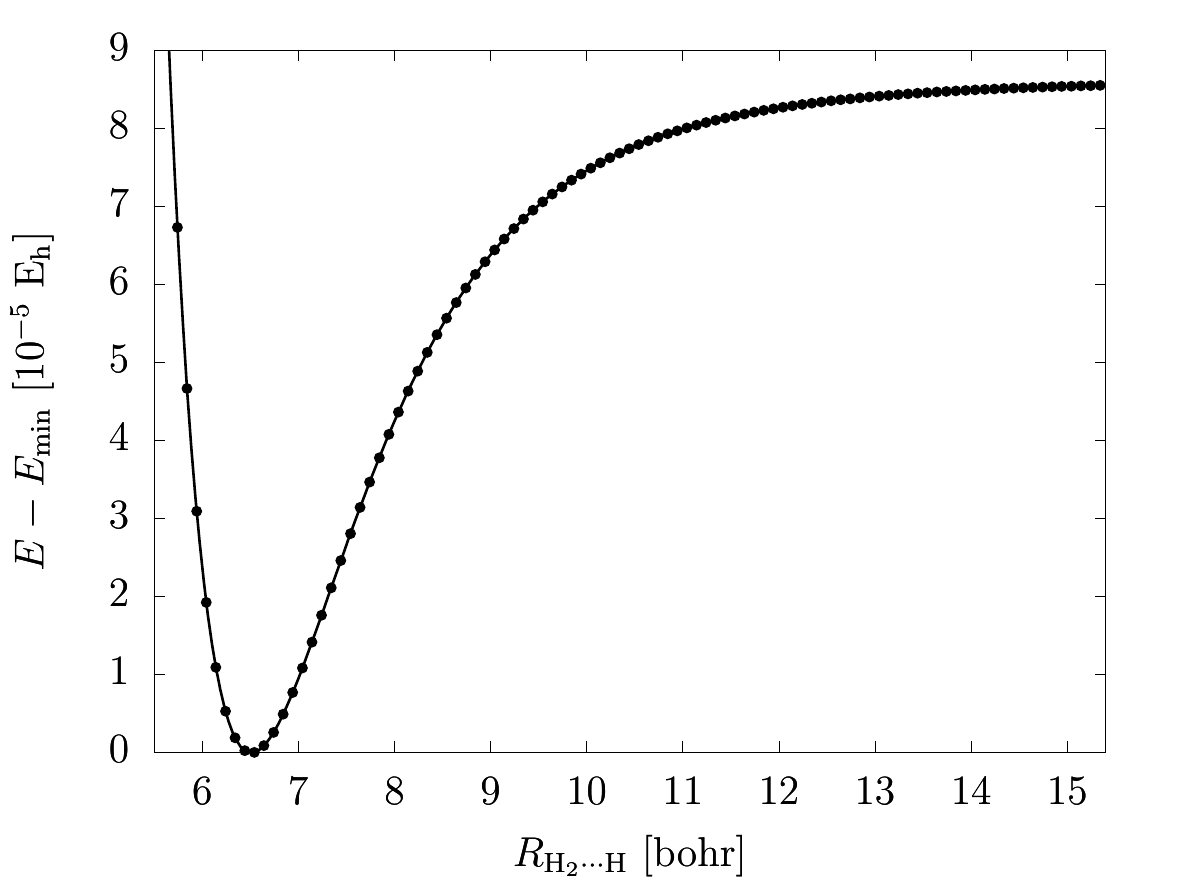} 
\caption{%
  Potential energy cut of the H$_3$ system converged in the present work with an estimated sub-ppm precision. Along the curve, the geometry of the H$_2$ unit is fixed at $R_{\text{H}_2}=1.4$~bohr.
  The lowest-energy datapoint corresponds to $E_\text{min}=-1.674\ 561\ 899\ \Eh$ and 
  $R_\text{min}=6.542$~bohr.
  \label{fig:pec}
}
\end{figure}

The energies (Fig.~\ref{fig:pec}) and optimized basis set parameters are deposited in the \som.

\section{Results and discussion\label{sec:res}}
We have carried out extensive single-point computations for the near-equilibrium geometry in the van-der-Waals well with $R\start_{\text{H}_2} = 1.4$~bohr and $R\start_{\text{H}_2\cdots \text{H}} = 6.442$~bohr first reported in Ref.~\cite{CafAdam2001}. This structure is close to the equilibrium geometry obtained with carefully conducted orbital-based computations \cite{MiGaPe-H3PES} (Table~\ref{tab:compare}). 
The energy of Ref.~\cite{CafAdam2001} computed with a small ECG basis is inaccurate, but later, large-scale computations were reported in Ref.~\cite{PaTuAd2009}.

At this geometry, the best energy obtained from the present work with 1200 ECGs (constructed by the initial fragment initialization, Sec.~\ref{sec:init}, followed by $n_\text{Powell}=3000$ Powell refinement cycles of the entire basis set) is $-1.674\ 561\ 687~\Eh$ (upper part of Table~\ref{tab:convergence}). Table~\ref{tab:convergence} also shows  the computed energy values for smaller basis sets that allow assessment of the convergence and extrapolation to the complete basis set (CBS) limit \cite{ConvergenceOfECG}. 

Direct comparison with Ref.~\cite{PaTuAd2009} requires further computation, because the extensively optimized energy reported in Ref.~\cite{PaTuAd2009} appears to belong to a 6.442~bohr distance of the hydrogen atom not from the center of nuclear mass of the H$_2$ unit, but from the closer proton of H$_2$. We think that this nuclear structure was used in Ref.~\cite{PaTuAd2009}, because we obtain good agreement for the energies when we perform the computation at this geometry, shown in the lower part of Table~\ref{tab:convergence}, corresponding to  
$R\start_{\text{H}_2}=1.40$~bohr and $\Rhh'=\Rhh\start+R\start_{\text{H}_2}/2=6.442\ \text{bohr}+0.700\ \text{bohr}=7.142$~bohr.

\begin{center}
\begin{table}
  \caption{%
  Convergence of the non-relativistic, ground-state energy of H$_3$ near the van-der-Waals equilibrium structure at $R_{\text{\Htwo{}}}=1.4$~bohr and $\Rhh = 6.442$~bohr
  taken from Ref.~\cite{CafAdam2001}.
  }
  \label{tab:convergence}
  \begin{tabular}{@{}c ll l@{}}
    \hline\hline\\[-0.35cm]
    \multicolumn{1}{c}{$\nb$}  &
    \multicolumn{1}{c}{Ansatz}  &
    \multicolumn{1}{c}{$n_\text{Powell}$}  &
    \multicolumn{1}{c}{$E\ [\Eh]$} \\
    \hline\\[-0.30cm]
    \multicolumn{4}{l}{\underline{$R_{\text{H}_2}$ = 1.40~bohr, $\Rhh=6.442$~bohr:} $^\text{a}$} \\[0.15cm]
    600  & $\lbrace\psi\hmol_{10n+i}\cdot \psi\hatom_i\rbrace$  & 2000  & $-1.674\ 560\ 470 $ \\
    800  & $\lbrace\psi\hmol_{10n+i}\cdot \psi\hatom_i\rbrace$  & 2000  & $-1.674\ 561\ 379 $ \\
    1000 & $\lbrace\psi\hmol_{10n+i}\cdot \psi\hatom_i\rbrace$  & 2000  & $-1.674\ 561\ 583 $ \\
    1200 &
      $\lbrace\psi\hmol_{10n+i}\cdot \psi\hatom_i\rbrace$ & 3000 & $-1.674\ 561\ 687 $ \\
    \multicolumn{3}{l}{[Extrapolation to $\nb\rightarrow\infty$:} & $-1.674\ 561\ 75(3)]$ \\
    \hline\hline\\[-0.30cm]
    \multicolumn{4}{l}{\underline{$R_{\text{H}_2}$ = 1.40~bohr, $\Rhh=7.142$~bohr:} $^\text{b}$} \\[0.15cm]
    1000 & \multicolumn{2}{l}{Ref.~\cite{PaTuAd2009}$^\text{c}$}  & $-1.674\ 547\ 421\ 00$ \\
    1200 & $\lbrace\psi\hmol_{10n+i}\cdot \psi\hatom_i\rbrace$  & 3000  &  $-1.674\ 547\ 750$ \\    
    \hline\hline
  \end{tabular}
  ~\\
    $^\text{a}$ %
    $R_{\text{H}_2}$ = 1.4~bohr, $\Rhh=6.442$~bohr, measured from the nuclear center of mass (NCM) of the H$_2$ unit.\\
    $^\text{b}$ %
    $R_{\text{H}_2}$ = 1.4~bohr, $\Rhh=7.142$~bohr (measured from the NCM of the H$_2$ unit), and corresponds to a 6.442~bohr distance measured from the nearer proton in the H$_2$ unit. \\
    $^\text{c}$ 
      Geometry $^\text{a}$ is claimed in Ref.~\cite{PaTuAd2009}, but it appears to be $^\text{b}$. The difference amounts to whether the distance of the hydrogen atom is measured from the NCM or the nearer proton.
\end{table}
\end{center}

We also note that the best energy value of Ref.~\cite{PaTuAd2009} computed in 6 months (using 12 CPU cores) was reproduced in this work (corresponding to the structure given in footnote $^\text{b}$ of Table~\ref{tab:convergence}) using the optimized fragment initialization technique (Sec.~\ref{sec:init}) followed by a few Powell refinement cycles in in 4 days. 
The computational benefit of the optimized fragment technique is significant in comparison with a computation \cite{PaTuAd2009} constructed from `scratch' immediately for the three-particle problem . 

Then, we continued the extensive refinement of the basis parameterization based on the variational principle, and the best result after 3 months computation (using 12 CPU cores) is reported in Table~\ref{tab:convergence}. The generation of the points along the PEC was started from this well-optimized parameterization by $\Delta\Rhh=\pm0.1$~bohr increments/decrements (running in parallel)
using the rescaling technique (Sec.~\ref{sec:init}), followed by 5 Powell refinement cycles at every step (before the next step was taken). The entire PEC generation took took 13 days using 12 CPU cores.

Finally, it is relevant to compare the ECG energies with the best orbital-based results underlying the CCI PES. For this reason, we have used a single rescaling step from the starting optimized parameterization (upper part of Table~\ref{tab:convergence}) to the $R\start_{\text{H}_2}=1.4$~bohr and $\Rhh\start=6.51205$~bohr structure, which was determined to be the global minimum structure at the MRCI/aug-cc-pV6Z level \cite{MiGaPe-H3PES}. The parameter rescaling, with a negligible computational cost, was followed by 5 Powell refinement cycles that took 4 hours.
Table~\ref{tab:compare} shows the energy values reported for the MRCI computations corresponding to the aug-cc-pVXZ (X = D, T, Q, 5, 6) and the `modified' correlation consistent aug-mcc-pVXZ (X = D, T, Q, 5, 6, 7) basis sets \cite{MiGaPe-H3PES,MiGaBrPe1999}. The ECG energy is already $74~\muEh$ lower, than the best MRCI value corresponding to the largest (aug-mcc-pV7Z) basis set.
Furthermore, we can confirm the estimated $\muEh$ precision of the CBS extrapolated energy from the mcc basis, whereas the extrapolated energy based on the regular correlation consistent basis is slightly lower than our current best estimate \cite{MiGaPe-H3PES}.

\begin{center}
\begin{table}[h]
  \caption{%
  Comparison of energies of various \emph{ab initio} computations.
  The equilibrium geometry, determined at the MRCI/aug-cc-pV6Z is $R_{\text{\Htwo{}}}=1.4015$~bohr and $\Rhh = 6.51205$~bohr \cite{MiGaPe-H3PES}.
  }
  \label{tab:compare}
  \begin{tabular}{@{}l l @{}}
    \hline\hline\\[-0.35cm]
    \multicolumn{1}{c}{Source}  &
    \multicolumn{1}{c}{$E$ [$\Eh$]}  \\
    \cline{1-2}  \\[-0.3cm]
    aug-cc-pVDZ $^\text{a}$    & $-1.664\ 339$ \\
    aug-cc-pVTZ $^\text{a}$    & $-1.672\ 540$ \\
    aug-cc-pVQZ $^\text{a}$    & $-1.673\ 902$ \\
    aug-cc-pV5Z $^\text{a}$    & $-1.674\ 332$ \\
    aug-cc-pV6Z $^\text{a}$    & $-1.674\ 445$ \\
    aug-mcc-pVTZ$^\text{a}$    & $-1.672\ 553$ \\
    aug-mcc-pVQZ$^\text{a}$    & $-1.673\ 917$ \\
    aug-mcc-pV5Z$^\text{a}$    & $-1.674\ 298$ \\
    aug-mcc-pV6Z$^\text{a}$    & $-1.674\ 430$ \\
    aug-mcc-pV7Z$^\text{a}$    & $-1.674\ 488$ \\ 
    MBE$^\text{cc}$(3,4 CBS)$^\text{b}$  & $-1.674\ 566$ \\
    MBE$^\text{mcc}$(6,7 CBS)$^\text{c}$ & $-1.674\ 562$ \\
    Present work ($\nb=1200$)$^\text{d}$ & $-1.674\ 562\ 264 $\\
    %
    \hline\hline
  \end{tabular}
   \begin{flushleft}
    $^\text{a}$ %
      Ref.~\cite{MiGaPe-H3PES}: MRCI energy. \\
    $^\text{b,\ c}$ %
      Ref.~\cite{MiGaPe-H3PES}: extrapolated CBS energy corresponding to
      the aug-cc-pV$X$Z ($X = 3, 4$) and aug-mcc-pV$X$Z ($X = 6, 7$) basis sets, respectively.
    \\
    $^\text{d}$ %
      Rescaled from the basis set optimized for the $(R\start_{\text{H}_2},R\start_{\text{H}_2\cdots\text{H}})$ structure in Table~\ref{tab:convergence}
      followed by 1000 Powell refinement cycles. 
  \end{flushleft} 
\end{table}
\end{center}

\section{Summary, conclusion, and outlook}
In summary, we have computed a benchmark-quality one-dimensional segment of the Born--Oppenheimer potential energy surface of the \Hthree{} system for a series of collinear nuclear configurations. The  electronic energies are estimated to be converged on the sub-parts-per-million level.

The depth of the van-der-Waals well 
was predicted to be $86(1)~\muEh$ at the $R_{\text{\Htwo{}}}=1.4015$~bohr and $\Rhh = 6.51205$~bohr geometry
in MRCI computations underlying the currently most precise potential energy surface of H$_3$ \cite{MiGaPe-H3PES}.
The variational computations reported in this work and using a (relatively small) explicitly correlated Gaussian basis set confirm this value and improve upon its precision by two orders of magnitude, $86.54(3)~\muEh$.
In order to achieve a similar precision for non-collinear nuclear structures, which have a lower order or no point-group symmetry, it will be necessary to use a larger basis set, which is certainly feasible. 

Regarding the broader context of this work, (non-)adiabatic perturbation theory \cite{Te2003,PaSpTe2007,PaKo2008,MaTe2019,MaFe2022} combined with leading-order relativistic and quantum electrodynamics (QED) corrections \cite{araki57,sucherPhD1958} are expected to provide a state-of-the-art theoretical description for this system. This framework has already been extensively used and tested for the lightest diatomic molecules \cite{KoPuCzLaPa2019,FeKoMa2020}. For the ground-electronic state of the H$_2$ molecule, the effect of the non-adiabatic-relativistic coupling has also been evaluated and was found to be non-negligible \cite{CzPuKoPa2018}. 
In this direction, the computation of a precise representation of the electronic wave function is a necessary first step that was demonstrated in this work to be feasible. The adiabatic, non-adiabatic and (regularized) relativistic and QED corrections can be evaluated at a couple of points using currently existing procedures \cite{MaFe2022,PaCeKo2005,FeKoMa2020,JeIrFeMa2022}.
At the same time, 
for a complete description of a polyatomic system like H$_3$,
these corrections must be computed over hundreds or thousands of nuclear configurations. 
This requires a fully automated evaluation and error control 
of all corrections, which may be especially challenging for the singular terms in the relativistic and QED expressions, and this requires further methodological and algorithmic developments that is left for future work.

\section{Acknowledgment}
Financial support of the European Research Council through a Starting Grant (No.~851421) is gratefully acknowledged. 
DF thanks a doctoral scholarship from
the ÚNKP-21-3 New National Excellence Program of the Ministry for Innovation and Technology from the source of the National Research, Development and Innovation Fund (ÚNKP-21-3-II-ELTE-41). 
We also thank Péter Jeszenszki for discussions about energy extrapolation for ECG basis sets.

 \bibliographystyle{elsarticle-num} 

\begin{thebibliography}{10}
\expandafter\ifx\csname url\endcsname\relax
  \def\url#1{\texttt{#1}}\fi
\expandafter\ifx\csname urlprefix\endcsname\relax\def\urlprefix{URL }\fi
\expandafter\ifx\csname href\endcsname\relax
  \def\href#1#2{#2} \def\path#1{#1}\fi

\bibitem{AoBaHe2005}
F.~J. Aoiz, L.~B. {N}ares, V.~J. Herrero, The {H+H}$_2$ reactive system.
  {P}rogress in the study of the dynamics of the simplest reaction, Int. Rev.
  Phys. Chem. 24~(1) (2005) 119--190.
\newblock \href {https://doi.org/10.1080/01442350500195659}
  {\path{doi:10.1080/01442350500195659}}.

\bibitem{Liu1973}
B.~Liu, Ab initio potential energy surface for linear {H}$_3$, J. Chem. Phys.
  58~(5) (1973) 1925--1937.
\newblock \href {https://doi.org/10.1063/1.1679454}
  {\path{doi:10.1063/1.1679454}}.

\bibitem{LSTH1}
P.~Siegbahn, B.~Liu, An accurate three‐dimensional potential energy surface
  for {H}$_3$, J. Chem. Phys. 68~(5) (1978) 2457--2465.
\newblock \href {https://doi.org/10.1063/1.436018}
  {\path{doi:10.1063/1.436018}}.

\bibitem{LSTH2}
D.~G. Truhlar, C.~J. Horowitz, Functional representation of {L}iu and
  {S}iegbahn’s accurate ab initio potential energy calculations for
  {H+H$_2$}, J. Chem. Phys. 68~(5) (1978) 2466--2476.
\newblock \href {https://doi.org/10.1063/1.436019}
  {\path{doi:10.1063/1.436019}}.

\bibitem{DMBE}
A.~J.~C. Varandas, F.~B. Brown, C.~A. Mead, D.~G. Truhlar, N.~C. Blais, A
  double many‐body expansion of the two lowest‐energy potential surfaces
  and nonadiabatic coupling for {H}$_3$, J. Chem. Phys. 86~(11) (1987)
  6258--6269.
\newblock \href {https://doi.org/10.1063/1.452463}
  {\path{doi:10.1063/1.452463}}.

\bibitem{BKMP}
A.~I. Boothroyd, W.~J. Keogh, P.~G. Martin, M.~R. Peterson, An improved {H}$_3$
  potential energy surface, J. Chem. Phys. 95~(6) (1991) 4343--4359.
\newblock \href {https://doi.org/10.1063/1.461758}
  {\path{doi:10.1063/1.461758}}.

\bibitem{BKMP2}
A.~I. Boothroyd, W.~J. Keogh, P.~G. Martin, M.~R. Peterson, A refined {H}$_3$
  potential energy surface, J. Chem. Phys. 104~(18) (1996) 7139--7152.
\newblock \href {https://doi.org/10.1063/1.471430}
  {\path{doi:10.1063/1.471430}}.

\bibitem{wu1999}
Y.-S. Wu, J.~Anderson, et~al., A very high accuracy potential energy surface
  for {H}$_3$, Phys. Chem. Chem. Phys. 1~(6) (1999) 929--937.
\newblock \href {https://doi.org/10.1039/A808797K}
  {\path{doi:10.1039/A808797K}}.

\bibitem{DiAn1994}
D.~L. Diedrich, J.~B. Anderson, Exact quantum {M}onte {C}arlo calculations of
  the potential energy surface for the reaction {H+ H$_2$ $\rightarrow$ H$_2$ +
  H}, J. Chem. Phys. 100~(11) (1994) 8089--8095.
\newblock \href {https://doi.org/10.1063/1.466802}
  {\path{doi:10.1063/1.466802}}.

\bibitem{BlombergLiu1985}
M.~R.~A. Blomberg, B.~Liu, The {H}$_3$ potential surface revisited, J. Chem.
  Phys. 82~(2) (1985) 1050--1051.
\newblock \href {https://doi.org/10.1063/1.448527}
  {\path{doi:10.1063/1.448527}}.

\bibitem{BaStLaPa1990}
C.~W. Bauschlicher, S.~R. Langhoff, H.~Partridge, A reevaluation of the {H}$_3$
  potential, Chem. Phys. Lett. 170~(4) (1990) 345--348.
\newblock \href {https://doi.org/10.1016/S0009-2614(90)87029-Q}
  {\path{doi:10.1016/S0009-2614(90)87029-Q}}.

\bibitem{PaBaStEu1993}
H.~Partridge, C.~W. Bauschlicher, J.~R. Stallcop, E.~Levin, Ab initio potential
  energy surface for {H–H$_2$}, J. Chem. Phys. 99~(8) (1993) 5951--5960.
\newblock \href {http://arxiv.org/abs/10.1063/1.465894}
  {\path{arXiv:10.1063/1.465894}}, \href {https://doi.org/10.1063/1.465894}
  {\path{doi:10.1063/1.465894}}.

\bibitem{DiAn1992}
D.~L. Diedrich, J.~B. Anderson, An accurate quantum {M}onte {C}arlo calculation
  of the barrier height for the reaction {H + H$_2$ $\rightarrow$ H$_2$ + H},
  Science 258~(5083) (1992) 786--788.
\newblock \href {https://doi.org/0.1126/science.258.5083.786}
  {\path{doi:0.1126/science.258.5083.786}}.

\bibitem{MiGaBrPe1999}
S.~L. Mielke, B.~C. Garrett, K.~A. Peterson, The utility of many-body
  decompositions for the accurate basis set extrapolation of ab initio data, J.
  Chem. Phys. 111~(9) (1999) 3806--3811.
\newblock \href {https://doi.org/10.1063/1.479683}
  {\path{doi:10.1063/1.479683}}.

\bibitem{RiAn2003}
K.~E. Riley, J.~B. Anderson, Higher accuracy quantum {M}onte {C}arlo
  calculations of the barrier for the {H}+{H}$_2$ reaction, J. Chem. Phys.
  118~(7) (2003) 3437--3438.
\newblock \href {https://doi.org/10.1063/1.1527012}
  {\path{doi:10.1063/1.1527012}}.

\bibitem{Hong-Xin2005}
H.-X. Huang, Exact {F}ixed-node {Q}uantum {M}onte {C}arlo: {D}ifferential
  {A}pproach, Chin. J. Chem. 23~(11) (2005) 1474--1478.
\newblock \href {https://doi.org/10.1002/cjoc.200591474}
  {\path{doi:10.1002/cjoc.200591474}}.

\bibitem{MiGaPe-H3PES}
S.~L. Mielke, B.~C. Garrett, K.~A. Peterson, A hierarchical family of global
  analytic {B}orn--{O}ppenheimer potential energy surfaces for the {H+H$_2$}
  reaction ranging in quality from double-zeta to the complete basis set limit,
  J. Chem. Phys. 116~(10) (2002) 4142--4161.
\newblock \href {https://doi.org/10.1063/1.1432319}
  {\path{doi:10.1063/1.1432319}}.

\bibitem{Mielke2003}
S.~L. Mielke, K.~A. Peterson, D.~W. Schwenke, B.~C. Garrett, D.~G. Truhlar,
  J.~V. Michael, M.-C. Su, J.~W. Sutherland, {$\mathrm{H}+{\mathrm{H}}_{2}$}
  {T}hermal {R}eaction: {A} {C}onvergence of {T}heory and {E}xperiment, Phys.
  Rev. Lett. 91 (2003) 063201.
\newblock \href {https://doi.org/10.1103/PhysRevLett.91.063201}
  {\path{doi:10.1103/PhysRevLett.91.063201}}.

\bibitem{CafAdam2001}
M.~Cafiero, L.~Adamowicz, Simultaneous optimization of molecular geometry and
  the wave function in a basis of {S}inger's n-electron explicitly correlated
  {G}aussians, Chem. Phys. Lett. 335~(5) (2001) 404--408.
\newblock \href {https://doi.org/10.1016/S0009-2614(01)00086-0}
  {\path{doi:10.1016/S0009-2614(01)00086-0}}.

\bibitem{PaTuAd2009}
M.~Pavanello, W.-C. Tung, L.~Adamowicz, How to calculate {H}$_3$ better, J.
  Chem. Phys. 131~(18) (2009) 184106.
\newblock \href {https://doi.org/10.1063/1.3257592}
  {\path{doi:10.1063/1.3257592}}.

\bibitem{SuzukiVarge1998}
Y.~Suzuki, K.~Varga, Stochastic {{Variational Approach}} to
  {{Quantum}}-{{Mechanical Few}}-{{Body Problems}}, Springer-Verlag, Berlin,
  Heidelberg, 1998.

\bibitem{CeKoPaSz2005}
W.~Cencek, J.~Komasa, K.~Pachucki, K.~Szalewicz, Relativistic {C}orrection to
  the {H}elium {D}imer {I}nteraction {E}nergy, Phys. Rev. Lett. 95 (2005)
  233004.
\newblock \href {https://doi.org/10.1103/PhysRevLett.95.233004}
  {\path{doi:10.1103/PhysRevLett.95.233004}}.

\bibitem{MaRe2012}
E.~M\'atyus, M.~Reiher, Molecular structure calculations: a unified quantum
  mechanical description of electrons and nuclei using explicitly correlated
  {G}aussian functions and the global vector representation, J. Chem. Phys. 137
  (2012) 024104.
\newblock \href {https://doi.org/10.1063/1.4731696}
  {\path{doi:10.1063/1.4731696}}.

\bibitem{Matyus2019}
E.~Mátyus, Pre-{Born}--{Oppenheimer} molecular structure theory, Mol. Phys.
  117~(5) (2019) 590--609.
\newblock \href {https://doi.org/10.1080/00268976.2018.1530461}
  {\path{doi:10.1080/00268976.2018.1530461}}.

\bibitem{Po2004}
M. J. D. Powell, The NEWUOA software for unconstrained optimization without
  derivatives (DAMTP 2004/NA05), Report no. NA2004/08,
  http://www.damtp.cam.ac.uk/user/na/reports04.html last accessed on January
  18, 2013.

\bibitem{Pachucki2010}
K.~Pachucki, Born--{O}ppenheimer potential for {${\mathrm{H}}_{2}$}, Phys. Rev.
  A 82 (2010) 032509.
\newblock \href {https://doi.org/10.1103/PhysRevA.82.032509}
  {\path{doi:10.1103/PhysRevA.82.032509}}.

\bibitem{PaAd2009}
M.~Pavanello, L.~Adamowicz, High-accuracy calculations of the ground, $1\
  ^1{A}_1'$, and the $2\ ^1{A}_1'$, $2\ ^3{A}_1'$, and $1\ ^1{E}'$ excited
  states of {H}$_3^+$, J. Chem. Phys. 130~(3) (2009) 034104.
\newblock \href {https://doi.org/10.1063/1.3058634}
  {\path{doi:10.1063/1.3058634}}.

\bibitem{KolosWolniewicz1964}
W.~Ko{\l}os, L.~Wolniewicz, Accurate {A}diabatic {T}reatment of the {G}round
  {S}tate of the {H}ydrogen {M}olecule, J. Chem. Phys. 41~(12) (1964)
  3663--3673.
\newblock \href {https://doi.org/10.1063/1.1725796}
  {\path{doi:10.1063/1.1725796}}.

\bibitem{CeKu1997}
W.~Cencek, W.~Kutzelnigg, Accurate adiabatic correction for the hydrogen
  molecule using the {B}orn--{H}andy formula, Chem. Phys. Lett. 266~(3-4)
  (1997) 383--387.
\newblock \href {https://doi.org/10.1016/S0009-2614(97)00017-1}
  {\path{doi:10.1016/S0009-2614(97)00017-1}}.

\bibitem{PaTuLeAd2009}
M.~Pavanello, W.-C. Tung, F.~Leonarski, L.~Adamowicz, New more accurate
  calculations of the ground state potential energy surface of {H}$_3^+$, J.
  Chem. Phys. 130~(7) (2009) 074105.
\newblock \href {https://doi.org/10.1063/1.3077193}
  {\path{doi:10.1063/1.3077193}}.

\bibitem{AdPa2012}
L.~Adamowicz, M.~Pavanello, Progress in calculating the potential energy
  surface of {H}$_3^+$, Philos. Trans. R. Soc. A 370~(1978) (2012) 5001--5013.
\newblock \href {https://doi.org/10.1098/rsta.2012.0101}
  {\path{doi:10.1098/rsta.2012.0101}}.

\bibitem{H3pPES2012}
M.~Pavanello, L.~Adamowicz, A.~Alijah, N.~F. Zobov, I.~I. Mizus, O.~L.
  Polyansky, J.~Tennyson, T.~Szidarovszky, A.~G. Császár,
  {C}alibration-quality adiabatic potential energy surfaces for {H}$_3^+$ and
  its isotopologues, J. Chem. Phys. 136~(18) (2012) 184303.
\newblock \href {https://doi.org/10.1063/1.4711756}
  {\path{doi:10.1063/1.4711756}}.

\bibitem{FeMa2019HH}
D.~Ferenc, E.~Mátyus, Non-adiabatic mass correction for excited states of
  molecular hydrogen: {Improvement} for the outer-well {$H\bar{H}$}
  {$^1\Sigma_g^+$} term values, J. Chem. Phys. 151~(9) (2019) 094101,
  publisher: American Institute of Physics.
\newblock \href {https://doi.org/10.1063/1.5109964}
  {\path{doi:10.1063/1.5109964}}.

\bibitem{FeMa2019EF}
D.~Ferenc, E.~Mátyus, Computation of rovibronic resonances of molecular
  hydrogen: {$EF\ ^1\Sigma_g^+$} inner-well rotational states, Phys. Rev. A
  100~(2) (2019) 020501.
\newblock \href {https://doi.org/10.1103/PhysRevA.100.020501}
  {\path{doi:10.1103/PhysRevA.100.020501}}.

\bibitem{FeKoMa2020}
D.~Ferenc, V.~I. Korobov, E.~M\'atyus, Nonadiabatic, {R}elativistic, and
  {L}eading-{O}rder {QED} {C}orrections for {R}ovibrational {I}ntervals of
  ${^{4}\mathrm{He}}_{2}^{+}$ (${X}\text{
  }{^{2}\mathrm{\ensuremath{\Sigma}}}_{u}^{+}$), Phys. Rev. Lett. 125 (2020)
  213001.
\newblock \href {https://doi.org/10.1103/PhysRevLett.125.213001}
  {\path{doi:10.1103/PhysRevLett.125.213001}}.

\bibitem{ConvergenceOfECG}
P.~Kopta, T.~Piontek, K.~Kurowski, M.~Puchalski, J.~Komasa, Convergence of
  {E}xplicitly {C}orrelated {G}aussian {W}ave {F}unctions, in: EScience on
  Distributed Computing Infrastructure - Volume 8500, Springer-Verlag, Berlin,
  Heidelberg, 2014, p. 459–474.
\newblock \href {https://doi.org/10.1007/978-3-319-10894-0_33}
  {\path{doi:10.1007/978-3-319-10894-0_33}}.

\bibitem{Te2003}
S.~Teufel, Adiabatic perturbation theory in quantum dynamics, Lecture Notes in
  Mathematics, Springer, 2003.

\bibitem{PaSpTe2007}
G.~Panati, H.~Spohn, S.~Teufel, The time-dependent {B}orn--{O}ppenheimer
  approximation, ESAIM: Math. Mod. Num. Anal. 41 (2007) 297.
\newblock \href {https://doi.org/10.1051/m2an:2007023}
  {\path{doi:10.1051/m2an:2007023}}.

\bibitem{PaKo2008}
K.~Pachucki, J.~Komasa, Nonadiabatic corrections to the wave function and
  energy, J. Chem. Phys. 129 (2008) 034102.
\newblock \href {https://doi.org/10.1063/1.2952517}
  {\path{doi:10.1063/1.2952517}}.

\bibitem{MaTe2019}
E.~M\'atyus, S.~Teufel, Effective non-adiabatic {H}amiltonians for the quantum
  nuclear motion over coupled electronic states, J. Chem. Phys. 151 (2019)
  014113.
\newblock \href {https://doi.org/10.1063/1.5097899}
  {\path{doi:10.1063/1.5097899}}.

\bibitem{MaFe2022}
E.~M\'atyus, D.~Ferenc, Vibronic mass computation for the
  {$EF$}--{$GK$}--{$HH$} $^1{\Sigma}_{\rm g}^+$ manifold of molecular hydrogen,
  Mol. Phys. (2022).

\bibitem{araki57}
H.~Araki, Quantum-electrodynamical corrections to energy-levels of helium,
  Prog. of Theor. Phys. 17 (1957) 619--642.
\newblock \href {https://doi.org/10.1143/PTP.17.619}
  {\path{doi:10.1143/PTP.17.619}}.

\bibitem{sucherPhD1958}
J.~Sucher, Energy levels of the two-electron atom, to order $\alpha^3$
  {Rydberg} ({C}olumbia {U}niversity) (1958).

\bibitem{KoPuCzLaPa2019}
J.~Komasa, M.~Puchalski, P.~Czachorowski, G.~\L{}ach, K.~Pachucki,
  Rovibrational energy levels of the hydrogen molecule through nonadiabatic
  perturbation theory, Phys. Rev. A 100 (2019) 032519.
\newblock \href {https://doi.org/10.1103/PhysRevA.100.032519}
  {\path{doi:10.1103/PhysRevA.100.032519}}.

\bibitem{CzPuKoPa2018}
P.~Czachorowski, M.~Puchalski, J.~Komasa, K.~Pachucki, Nonadiabatic
  relativistic correction in {${\mathrm{H}}_{2}$}, {${\mathrm{D}}_{2}$}, and
  {HD}, Phys. Rev. A 98 (2018) 052506.
\newblock \href {https://doi.org/10.1103/PhysRevA.98.052506}
  {\path{doi:10.1103/PhysRevA.98.052506}}.

\bibitem{PaCeKo2005}
K.~Pachucki, W.~Cencek, J.~Komasa, On the acceleration of the convergence of
  singular operators in {Gaussian} basis sets, J. Chem. Phys. 122~(18) (2005)
  184101.
\newblock \href {https://doi.org/10.1063/1.1888572}
  {\path{doi:10.1063/1.1888572}}.

\bibitem{JeIrFeMa2022}
P.~Jeszenszki, R.~T. Ireland, D.~Ferenc, E.~M{\'a}tyus, On the inclusion of
  cusp effects in expectation values with explicitly correlated {{Gaussians}},
  Int. J. Quant. Chem. (2021).
\newblock \href {https://doi.org/doi.org/10.1002/qua.26819}
  {\path{doi:doi.org/10.1002/qua.26819}}.

\end{thebibliography}

\clearpage

\onecolumn

\section*{Supplementary Material}


\begin{table}[h]
  \caption{%
  Potential energy curve of collinear \Hthree{}, with collinear hydrogen atoms and the H$_2$ structure fixed at $R_{\text{H}_2}=1.4$~bohr.
  }
  \label{tab:pec}
  \begin{tabular}{@{}c c  c c  c c @{}}
    \hline\hline\\[-0.35cm]
    \multicolumn{1}{c}{$\Rhh$ [bohr]}  &
    \multicolumn{1}{c}{$E$ [$\Eh$]}  &
    \multicolumn{1}{c}{$\Rhh$ [bohr]}  &
    \multicolumn{1}{c}{$E$ [$\Eh$]}  &
    \multicolumn{1}{c}{$\Rhh$ [bohr]}  &
    \multicolumn{1}{c}{$E$ [$\Eh$]}  \\
    \cline{1-6}  \\[-0.3cm]
    3.942  &	$-1.671\ 027\ 070$	&	7.742  &	$-1.674\ 527\ 262$	&	11.542 &	$-1.674\ 480\ 330$	\\
    4.042  &	$-1.671\ 542\ 481$	&	7.842  &	$-1.674\ 524\ 133$	&	11.642 &	$-1.674\ 480\ 079$	\\
    4.142  &	$-1.671\ 995\ 736$	&	7.942  &	$-1.674\ 521\ 142$	&	11.742 &	$-1.674\ 479\ 843$	\\
    4.242  &	$-1.672\ 392\ 125$	&	8.042  &	$-1.674\ 518\ 294$	&	11.842 &	$-1.674\ 479\ 622$	\\
    4.342  &	$-1.672\ 736\ 967$	&	8.142  &	$-1.674\ 515\ 593$	&	11.942 &	$-1.674\ 479\ 415$	\\
    4.442  &	$-1.673\ 035\ 388$	&	8.242  &	$-1.674\ 513\ 039$	&	12.042 &	$-1.674\ 479\ 220$	\\
    4.542  &	$-1.673\ 292\ 312$	&	8.342  &	$-1.674\ 510\ 633$	&	12.142 &	$-1.674\ 479\ 036$	\\
    4.642  &	$-1.673\ 512\ 388$	&	8.442  &	$-1.674\ 508\ 369$	&	12.242 &	$-1.674\ 478\ 864$	\\
    4.742  &	$-1.673\ 699\ 926$	&	8.542  &	$-1.674\ 506\ 245$	&	12.342 &	$-1.674\ 478\ 702$	\\
    4.842  &	$-1.673\ 858\ 896$	&	8.642  &	$-1.674\ 504\ 254$	&	12.442 &	$-1.674\ 478\ 549$	\\
    4.942  &	$-1.673\ 992\ 934$	&	8.742  &	$-1.674\ 502\ 391$	&	12.542 &	$-1.674\ 478\ 405$	\\
    5.042  &	$-1.674\ 105\ 325$	&	8.842  &	$-1.674\ 500\ 651$	&	12.642 &	$-1.674\ 478\ 269$	\\
    5.142  &	$-1.674\ 198\ 999$	&	8.942  &	$-1.674\ 499\ 026$	&	12.742 &	$-1.674\ 478\ 141$	\\
    5.242  &	$-1.674\ 276\ 566$	&	9.042  &	$-1.674\ 497\ 510$	&	12.842 &	$-1.674\ 478\ 020$	\\
    5.342  &	$-1.674\ 340\ 341$	&	9.142  &	$-1.674\ 496\ 096$	&	12.942 &	$-1.674\ 477\ 906$	\\
    5.442  &	$-1.674\ 392\ 365$	&	9.242  &	$-1.674\ 494\ 779$	&	13.042 &	$-1.674\ 477\ 799$	\\
    5.542  &	$-1.674\ 434\ 419$	&	9.342  &	$-1.674\ 493\ 552$	&	13.142 &	$-1.674\ 477\ 697$	\\
    5.642  &	$-1.674\ 468\ 056$	&	9.442  &	$-1.674\ 492\ 409$	&	13.242 &	$-1.674\ 477\ 601$	\\
    5.742  &	$-1.674\ 494\ 619$	&	9.542  &	$-1.674\ 491\ 345$	&	13.342 &	$-1.674\ 477\ 510$	\\
    5.842  &	$-1.674\ 515\ 270$	&	9.642  &	$-1.674\ 490\ 355$	&	13.442 &	$-1.674\ 477\ 424$	\\
    5.942  &	$-1.674\ 531\ 005$	&	9.742  &	$-1.674\ 489\ 433$	&	13.542 &	$-1.674\ 477\ 343$	\\
    6.042  &	$-1.674\ 542\ 675$	&	9.842  &	$-1.674\ 488\ 574$	&	13.642 &	$-1.674\ 477\ 266$	\\
    6.142  &	$-1.674\ 551\ 007$	&	9.942  &	$-1.674\ 487\ 775$	&	13.742 &	$-1.674\ 477\ 193$	\\
    6.242  &	$-1.674\ 556\ 616$	&	10.042 &	$-1.674\ 487\ 030$	&	13.842 &	$-1.674\ 477\ 124$	\\
    6.342  &	$-1.674\ 560\ 021$	&	10.142 &	$-1.674\ 486\ 337$	&	13.942 &	$-1.674\ 477\ 058$	\\
    6.442  &	$-1.674\ 561\ 676$	&	10.242 &	$-1.674\ 485\ 691$	&	14.042 &	$-1.674\ 476\ 996$	\\
    6.542  &	$-1.674\ 561\ 899$	&	10.342 &	$-1.674\ 485\ 088$	&	14.142 &	$-1.674\ 476\ 937$	\\
    6.642  &	$-1.674\ 561\ 042$	&	10.442 &	$-1.674\ 484\ 527$	&	14.242 &	$-1.674\ 476\ 880$	\\
    6.742  &	$-1.674\ 559\ 343$	&	10.542 &	$-1.674\ 484\ 003$	&	14.342 &	$-1.674\ 476\ 827$	\\
    6.842  &	$-1.674\ 557\ 010$	&	10.642 &	$-1.674\ 483\ 515$	&	14.442 &	$-1.674\ 476\ 777$	\\
    6.942  &	$-1.674\ 554\ 214$	&	10.742 &	$-1.674\ 483\ 058$	&	14.542 &	$-1.674\ 476\ 728$	\\
    7.042  &	$-1.674\ 551\ 095$	&	10.842 &	$-1.674\ 482\ 632$	&	14.642 &	$-1.674\ 476\ 683$	\\
    7.142  &	$-1.674\ 547\ 764$	&	10.942 &	$-1.674\ 482\ 234$	&	14.742 &	$-1.674\ 476\ 639$	\\
    7.242  &	$-1.674\ 544\ 313$	&	11.042 &	$-1.674\ 481\ 862$	&	14.842 &	$-1.674\ 476\ 598$	\\
    7.342  &	$-1.674\ 540\ 812$	&	11.142 &	$-1.674\ 481\ 514$	&	14.942 &	$-1.674\ 476\ 559$	\\
    7.442  &	$-1.674\ 537\ 318$	&	11.242 &	$-1.674\ 481\ 188$	&	15.042 &	$-1.674\ 476\ 521$	\\
    7.542  &	$-1.674\ 533\ 875$	&	11.342 &	$-1.674\ 480\ 883$	&	15.142 &	$-1.674\ 476\ 485$	\\
    7.642  &	$-1.674\ 530\ 514$	&	11.442 &	$-1.674\ 480\ 598$	&	15.242 &	$-1.674\ 476\ 451$	\\
    \hline\hline
  \end{tabular}
\end{table}

\end{document}